\documentstyle[12pt]{article}

\title{Dimensional transmutation and symmetry breaking
in Maxwell-Chern-Simons scalar QED}

\author{F. S. Nogueira and N. F. Svaiter \\
Centro Brasileiro de Pesquisas Fisicas - CBPF \\
Rua Dr. Xavier Sigaud 150, Rio de Janeiro, RJ 22290-180 \\
BRAZIL}

\begin{document} \setcounter{equation}{0}

\maketitle

\begin{abstract}
The mechanism of dimensional transmutation is discussed in
the context of Maxwell-Chern-Simons scalar QED. The method
used is non-perturbative. The effective
potential describes a broken symmetry state. It is found
that the symmetry breaking vacuum is more stable when the
Chern-Simons mass is different from zero.

\end{abstract}

Pacs numbers: 11.10.Ef, 11.10.Gh

\newpage

\section{Introduction} \setcounter{equation}{0}

The idea of spontaneous symmetry breaking allows one to
describe a wide class of phenomena in both condensed matter and
particle physics. In condensed matter physics it furnishes a good
phenomenological description of many interesting phenomena. For
instance, the magnetic materials and superconduncting
compounds are among the physical systems susccessfully described
by the broken symmetry picture \cite{Anderson}. The case of
superconductivity is of particular interest since it involves a
local gauge symmetry. Indeed, the idea of spontaneously broken
gauge theories have its origin in superconductivity theory
\cite{Anderson1}.
Phenomenologically, it is well described by a Landau-Ginsburg
action with the charged scalar field minimally coupled to
Abelian gauge fields. This is just the action of scalar QED. In
order to describe a broken symmetry solution, an imaginary mass
is attributed to the scalar field. Thus, the gauge field acquire
mass through the Higgs mechanism \cite{Higgs}. This is the origin
of the well known Meissner effect. The Higgs mechanism was shown
to give also a good description of particle physics phenomenology
through the so called standard model of elementary particles. In
fact, it is on the basis of the experimentally tested electroweak
theory. According to the standard model, the Higgs mechanism is
responsible for the masses of the vector bosons, the $W^{\pm}$ and
$Z^{0}$.

Another path to a broken symmetry state is the
Coleman-Weinberg mechanism \cite{Coleman}. The Coleman-Weinberg
mechanism consists in induce the symmetry breaking via
radiative corrections. Thus, it is not attributed an
imaginary mass to the scalar particle. Instead, the
renormalized mass is zero and the symmetry breaking is not
manifest at the tree level as is the case of the Higgs
mechanism. The Coleman-Weinberg mechanism is very useful in
the construction of grand unified theories and in cosmological
models \cite{Guth}. A very interesting feature of this method is
the phenomenon of dimensional transmutation. Dimensional
transmutation occurs as a consequence of the breaking of the
symmetry. It consists of a reduction in the number of dimensionless
couplings which are replaced by corresponding dimensionful
parameters. For example, in the case of massless scalar $(QED)_{4}$
the theory has two dimensionless couplings, namely, $e^{2}$ and
$\lambda$. Here, $e^{2}$ correspond to the electromagnetic coupling
while $\lambda$ is the coupling of the scalar particle
self-interaction. The symmetry breaking through the Coleman-Weinberg
mechanism allows the elimination of the scalar self-coupling in
favour of the electromagnetic coupling. This is done at the
price of the
introduction of a dimensionful parameter in replacement of
$\lambda$. It turns out that this parameter
corresponds to the vacuum expectation
value of the scalar field. Thus, we have just two parameters as
before but only one is dimensionless.

In order to the phenomenon of dimensional transmutation takes
place it is necessary the presence of at least one
dimensionless parameter in the theory. This means that we cannot
obtain a similar situation in scalar $(QED)_{3}$ if we
restrict ourselves to a $(\phi^{\dag}\phi)^{2}$
interaction in the scalar
sector. Therefore, it is necessary the inclusion of a
$(\phi^{\dag}\phi)^{3}$ interaction. The resulting coupling will
be dimensionless and the
theory is renormalizable rather than super-renormalizable.
Another important point concerning tridimensional
QED is that it admits the inclusion of a
Chern-Simons term \cite{Deser}.
With all these terms collected we can build the
more general renormalizable scalar QED in $d=3$.

On the basis of the above discussion we can legitimately ask the
following question. Is it possible to implement the Coleman-Weinberg
mechanim in a renormalizable Maxwell-Chern-Simons QED? It is the
main aim of this paper to show that the answer to this question
is affirmative. In order to achieve this goal we will use a simple
non-perturbative approach. Recently the Coleman-Weinberg mechanism in
massless scalar $(QED)_{4}$ was studied non-perturbatively
\cite{Malbouisson}. It is shown
that the one-loop result of
Coleman and Weinberg can be established beyond the range of
validity of perturbation theory. For defineteness we will
rederive briefly the Coleman-Weinberg effective potential in
the approach of ref.\cite{Malbouisson}. This will help to fix the
ideas and will be done in section 2. In section 3 we use the
method of section 2 to obtain the effective potential for the
Maxwell-Chern-Simons scalar QED. We use the more general
renormalizable action which means that a $(\phi^{\dag}\phi)^{3}$
interaction term is included. We establish then the symmetry
breaking and dimensional transmutation in the massless case.
Finally, we discuss the results in section 4. In this paper we
use $\hbar=c=1$.

\section{Warm up: scalar $(QED)_{4}$} \setcounter{equation}{0}

The aim of this section is mainly to introduce the method that
will be used in the next section. The essence of the method is
that it is almost a tree level manipulation. It is worth to
point out that the Higgs mechanism works at the tree level if
one chooses the unitary gauge. In this case symmetry breaking
is manifest at tree level. We will make almost the same thing
with respect to the Coleman-Weinberg mechanism. This is
achieved by noting that in an Abelian theory the action is
quadratic in the gauge fields allowing a straightforward
Gaussian integration. This nice feature is also a weakness of
our method since it is not possible to generalize the
procedure to non-Abelian gauge fields. Just like in the case of
the Higgs mechanism we will find convenient to work in the unitary
gauge. The unitary gauge parametrization is obtained by
integrating out exactly the gauge freedom. The scalar sector is
then rewritten in terms of the real scalar field
$\rho(x)$ where $\rho^{2}=\phi_{1}^{2}+\phi_{2}^{2}$ where
$\phi_{1}$ and $\phi_{2}$ are respectively the real and
imaginary parts of the field $\phi$. Thus, after
straightforward integration of the gauge fields one obtains the
following Euclidean effective action \cite{Malbouisson}:

\begin{eqnarray}
\label{2.1}
S_{eff}[\rho]&=&\frac{1}{2}\ln\det[\delta_{\mu\nu}(-\Box+e^{2}\rho^{
2})+\partial_{\mu}\partial_{\nu}]-\delta^{4}(0){\int}d^{4}x\ln(e
\rho) \nonumber \\
&&+{\int}d^{4}x\left[\frac{1}{2}\rho(-\Box+m^{2})\rho+\frac{
\lambda}{4!}\rho^{4}\right].
\end{eqnarray}
This is an exact expression. The factor
$-\delta^{4}(0){\int}d^{4}x\ln(e\rho)$ arises from the
exponentiation of the Jacobian $\det(e\rho)$ which occurs
in the functional measure as a result of the unitary gauge
parametrization. Looking for a constant saddle point
$<\rho>$ we find

\begin{equation}
\label{2.2}
{\int}d^{4}x\left(m^{2}<\rho>+\frac{\lambda}{3!}<\rho>^{3}
-\frac{\delta^{4}(0)}{<\rho>}\right)+e^{2}<\rho>Tr
D_{\mu\nu}(x-x')=0,
\end{equation}
where $D_{\mu\nu}(x-x')$ is the propagator of the massive vector
field with mass $e^{2}<\rho>^{2}$. By evaluating explicitly the
trace of the propagator we obtain an exact cancellament of the
divergent factor proportional to $\delta^{4}(0)$. The solution
to Eq.(\ref{2.2}) that will concern us consists of the gap
equation:

\begin{equation}
\label{2.3}
<\rho>^{2}=-\frac{6m^{2}}{\lambda}-\frac{18e^{2}}{\lambda}
\int\frac{d^{4}p}{(2\pi)^{4}}\frac{1}{p^{2}+e^{2}<\rho>^{2}}.
\end{equation}
After evaluation of the integral above using a cutoff $\Lambda$
the gap equation becomes

\begin{equation}
\label{2.4}
\lambda_{R}<\rho>^{2}=-6m_{R}^{2}-\frac{9e^{4}}{8\pi^{2}}
<\rho>^{2}\ln\frac{e^{2}<\rho>^{2}}{\mu^{2}},
\end{equation}
where we have defined the renormalized parameters:

\begin{equation}
\label{2.5}
m_{R}^{2}=m^{2}+\frac{3e^{2}}{16\pi^{2}}\Lambda^{2},
\end{equation}

\begin{equation}
\label{2.6}
\lambda_{R}=\lambda+\frac{9e^{4}}{8\pi^{2}}\ln\frac{\mu^{2}}{
\Lambda^{2}}.
\end{equation}
In above $\mu$ is an arbitrary renormalzation scale. Now we
demand that all stationary points are solutions to Eq.(\ref{2.4}).
A broken symmetry solution corresponds to a local maximum at
the origin and two degenerate absolute minima of the
effective potential. The solution
corresponding to the local maximum $<\rho>_{max}=0$ is a
solution to Eq.(\ref{2.3}) only if $m_{R}^{2}=0$. Now, let the
solution corresponding to the minimum be given by
$<\rho>_{min}=\sigma$. We obtain that

\begin{equation}
\label{2.7}
\lambda_{R}=-\frac{9e^{4}}{8\pi^{2}}\ln\frac{e^{2}\sigma^{2}}{
\mu^{2}}.
\end{equation}
By considering a $x$-independent background field
$\overline{\rho}$ in the expression for the effective action,
Eq.(\ref{2.1}), and evaluating explicitly the
logarithm of the determinant we obtain the following
expression for the effective potential:

\begin{equation}
\label{2.8}
V(\overline{\rho})=\frac{\lambda_{R}}{24}\overline{\rho}^{4}+
\frac{3e^{4}}{64\pi^{2}}\overline{\rho}^{4}
\ln\frac{e^{2}\overline{\rho}^{2}}{
\mu^{2}}-\frac{3e^{4}}{128\pi^{2}}\overline{\rho}^{4},
\end{equation}
where we have assumed that $m_{R}^{2}=0$. Substituting
Eq.(\ref{2.7}) in Eq.(\ref{2.8}) one obtains

\begin{equation}
\label{2.9}
V(\overline{\rho})=\frac{3e^{4}}{64\pi^{2}}\overline{\rho}^{4}
\left(\ln\frac{\overline{\rho}^{2}}{\sigma^{2}}-\frac{1}{2}
\right).
\end{equation}
Eq.(\ref{2.9}) is just the Coleman-Weinberg potential
\cite{Coleman}. It is important to stress that we established
non-perturbatively the one-loop result of ref.\cite{Coleman}.
With the usual perturbative scheme it is assumed that
$\lambda_{R}{\sim}e^{4}$ and corrections ${\sim}\lambda_{R}^{2}$
are neglected. This assumption is not necessary here. Note
that we rederive the Coleman-Weinberg result in a mean field
like approximation with the help of a gap equation. The result is
obtained as the 'tree level` of the effective action,
Eq.(\ref{2.1}). Note also that the derivation given here is not
restricted to a small value of $e^{2}$.

\section{Maxwell-Chern-Simons scalar QED} \setcounter{equation}{0}

In this section we use the method of the previous section
applied to the case of the Maxwell-Chern-Simons scalar QED.
We work in the unitary gauge as in the last section. The
Euclidean effective action resulting
from the exact integration of the
vector fields is given by

\begin{eqnarray}
\label{3.1}
S_{eff}^{CS}[\rho]&=&\frac{1}{2}\ln\det[\delta_{\mu\nu}(
-\Box+e^{2}\rho^{2})+\partial_{\mu}\partial_{\nu}+i\theta
\epsilon_{\mu\lambda\nu}\partial_{\lambda}]-\delta^{3}(0)
{\int}d^{3}x\ln(e\rho) \nonumber \\
&&+{\int}d^{3}x\left[\frac{1}{2}\rho(-\Box+m^{2})\rho+
\frac{\lambda}{4!}\rho^{4}+\frac{\eta}{6!}\rho^{6}\right].
\end{eqnarray}
In above, $\theta$ is the Chern-Simons mass and $\eta$ is a
dimensionless coupling. Note that in $d=3$ the parameters
$e^{2}$ and $\lambda$ have dimension of mass. Eq.(\ref{3.1})
corresponds to the more general renormalizable action. Without
the term $\rho^{6}$ the action would be super-renormalizable and
we would have only dimensionful couplings. The presence of at
least one dimensionless coupling is crucial for the
phenomenon of dimensional transmutation. Otherwise, there is
nothing to transmute.

As before, we look for a constant saddle point solution to the
effective action. Then, stationarity of the action with respect
to this saddle point implies the following gap equation:

\begin{eqnarray}
\label{3.2}
e^{2}\left(1+\frac{|\theta|}{\sqrt{\theta^{2}+4e^{2}<\rho>^{2}}}
\right)\int\frac{d^{3}p}{(2\pi)^{3}}\frac{1}{p^{2}+
M_{+}^{2}(<\rho>^{2})} \nonumber \\
+e^{2}\left(1-\frac{|\theta|}{\sqrt{\theta^{2}+4e^{2}<\rho>^{2}}}
\right)\int\frac{d^{3}p}{(2\pi)^{3}}\frac{1}{p^{2}+
M_{-}^{2}(<\rho>^{2})} \nonumber \\
+m^{2}+\frac{\lambda}{6}<\rho>^{2}+\frac{\eta}{120}
<\rho>^{4}=0,
\end{eqnarray}
where the $M_{\pm}^{2}$ are defined by

\begin{equation}
\label{3.3}
M_{\pm}^{2}(<\rho>^{2})=e^{2}<\rho>^{2}+\frac{\theta^{2}}{2}
\pm\frac{|\theta|}{2}\sqrt{\theta^{2}+4e^{2}<\rho>^{2}}.
\end{equation}
By using an ultraviolet cutoff $\Lambda$ it is straightforward
to compute the integrals in Eq.(\ref{3.2}). The gap equation
becomes

\begin{eqnarray}
\label{3.4}
-\frac{e^{2}}{4\pi}|M_{+}(<\rho>^{2})|\left(1+\frac{
|\theta|}{\sqrt{\theta^{2}+4e^{2}<\rho>^{2}}}\right) \nonumber \\
-\frac{e^{2}}{4\pi}|M_{-}(<\rho>^{2})|\left(1-\frac{|\theta|}{
\sqrt{\theta^{2}+4e^{2}<\rho>^{2}}}\right) \nonumber \\
+\frac{\lambda}{6}
<\rho>^{2}+\frac{\eta}{120}<\rho>^{4}=0.
\end{eqnarray}
We have assumed, just as in the previous section, that the
renormalized mass is zero.

The effective potential is obtained from Eq.(\ref{3.1}) and is
given by

\begin{equation}
\label{3.5}
V(\overline{\rho})=-\frac{1}{12\pi}[|M_{+}(\overline{\rho}^{2})|^{3}
+|M_{-}(\overline{\rho}^{2})|^{3}]+\frac{\lambda}{4!}
\overline{\rho}^{4}+\frac{\eta}{6!}\overline{\rho}^{6}.
\end{equation}
Note that, in contrast to the calculations performed in the
previous section, no renormalization scale arises here. This is a
special feature of the $d=3$ case.

Let us consider the solution to the gap equation associated to
the symmetry breaking
minimum of the potential, $<\rho>_{min}=\sigma$. Solving
Eq.(\ref{3.4}) for $\eta$ and substituting the result in
Eq.(\ref{3.5}) we get

\begin{eqnarray}
\label{3.6}
V(\overline{\rho})&=&-\frac{1}{12\pi}[|M_{+}(\overline{
\rho}^{2})|^{3}+|M_{-}(\overline{\rho}^{2})|^{3}]+\frac{\lambda}{24}
\overline{\rho}^{4} \nonumber \\
&&+\frac{1}{12\sigma^{2}}\left\{\frac{e^{2}}{2\pi\sigma^{2}}
\left[|M_{+}(\sigma^{2})|\left(1+\frac{|\theta|}{
\sqrt{\theta^{2}+4e^{2}\sigma^{2}}}\right)
\right.\right. \nonumber \\
&&\left.\left.+|M_{-}(\sigma^{2})|
\left(1-\frac{|\theta|}{\sqrt{\theta^{2}
+4e^{2}\sigma^{2}}}\right)\right]-\frac{\lambda}{3}
\right\}\overline{\rho}^{6}.
\end{eqnarray}
Note that dimensional transmutation has occurred. We had before
four parameters, $\theta$,
$e^{2}$, $\lambda$ and $\eta$, the parameter
$\eta$ being dimensionless. Now, we remain with the same number of
parameters but the dimensionless parameter $\eta$ has disappeared
and has been replaced by the parameter $\sigma^{2}$, which has the
dimension of mass.

The above effective potential corresponds to a broken symmetry
phase with one local maximum at $<\rho>_{max}=0$ and with two
degenerate absolute minima, at $\pm\sigma$. If $\lambda>0$ we have
that the stability condition $\lambda\leq\lambda_{c}$ must holds,
where the critical parameter $\lambda_{c}$ is given by

\begin{eqnarray}
\label{3.7}
\lambda_{c}&=&\frac{3e^{2}}{2\pi\sigma^{2}}\left[
|M_{+}(\sigma^{2})|\left(1+\frac{|\theta|}{\sqrt{\theta^{2}+
4e^{2}\sigma^{2}}}\right)
\right. \nonumber \\
&&\left.
+|M_{-}(\sigma^{2})|\left(1-\frac{|\theta|}{\sqrt{
\theta^{2}+4e^{2}\sigma^{2}}}\right)\right].
\end{eqnarray}
If $\lambda>\lambda_{c}$ the potential is unbounded below and
the vacuum is unstable. However, if $\lambda<0$ no
stability condition is necessary because the potential will
be always be bounded from below. instability with respect to the
value of $\eta$, rather than $\lambda$, was considered
non-perturbatively in ref.\cite{Moshe} in the context of a
$O(N)$ symmetric $\phi^{6}_{3}$ theory. It was found that
the vacuum is unstable for $\eta>\eta_{c}$, $\eta_{c}$ being
the critical value of $\eta$. Here we have a similar situation.
The vacuum is unstable for $\lambda>\lambda_{c}$.
The stability condition
(\ref{3.7}) implies that in the pure scalar limit
$e^{2}=\theta=0$ we must have necessarily $\lambda{\leq}0$. The
limit $\lambda=\lambda_{c}$ corresponds to a
Maxwell-Chern-Simons theory without the term $(\phi^{\dag}\phi)^{3}$.
This situation with $\theta=0$ was already studied from a
perturbative point of view \cite{Appelquist}. The authors of
ref.\cite{Appelquist} argued that the symmetry
breaking obtained by the one-loop result
is spurious. Their calculations were also performed at the
critical point $m_{R}^{2}=0$. In this case the radiative corrections
failed in induce the symmetry breaking because of the absence of
the $(\phi^{\dag}\phi)^{3}$ term and,
therefore, of a dimensionless coupling.
Here we have made a pseudo-tree-level
analysis and, therefore, our result correspond to the leading
contribution to the effective potential. For this
reason, we argue that the fluctuations cannot turn the
asymmetric phase we found into a symmetric one. Our picture of the
Coleman-Weinberg mechanism is similar to the case of the
Higgs mechanism where formal manipulations performed at the tree
level in the unitary gauge produce a broken symmetry state. Since
the Coleman-Weinberg mechanism is based on the idea that
quantum fluctuations may induce symmetry breaking, we must perform
the tree level analysis in an effective field theory, described
in the case $d=4$ by the effective action, Eq.(\ref{2.1}), and in
the case $d=3$ by the effective action, Eq.(\ref{3.1}).

A further feature of the effective potential given by Eq.(\ref{3.6})
is that $V(0)=0$ only if $\theta=0$. If $\theta{\neq}0$ we have that
$V(0)<0$. The scalar field $\rho$ have the same vacuum
expectation value for all values of $\theta$. However,
the vacuum energy is lower in the case $\theta{\neq}0$ than in the
case $\theta=0$. This means that the renormalizable
(that is, with the term $(\phi^{\dag}\phi)^{3}$
included) Maxwell-Chern-Simons scalar QED is more stable than the
non-topological scalar QED in $d=3$.

Let us compare our results with the one-loop calculation. The
one-loop effective potential is given by

\begin{eqnarray}
\label{3.8}
V(\overline{\rho})&=&-\frac{1}{12\pi}[M_{+}^{3}(\overline{\rho}^{2})
+M_{-}^{3}(\overline{\rho}^{2})]-\frac{1}{12\pi}\left(\frac{
\lambda}{2}\overline{\rho}^{2}+\frac{\eta}{24}\overline{\rho}^{4}
\right)^{3/2} \nonumber \\
&&+\frac{\lambda}{4!}\overline{\rho}^{4}+\frac{\eta}{6!}
\overline{\rho}^{6}.
\end{eqnarray}
Here, $\eta$ is an independent parameter, that is, it is not
written in terms of the other parameters. When $\eta=\theta=0$,
Eq.(\ref{3.8}) agrees with
ref.\cite{Appelquist}. The situation which $\eta=0$ but
$\theta{\neq}0$ corresponds in our case to
$\lambda=\lambda_{c}$.
In this case the one-loop result have a term
which is absent in our result, a term proportional to
$\lambda^{3/2}|\overline{\rho}|^{3}$. However, in the loop
expansion given here $\lambda$ is not written as a function of
other parameters of the theory, as in Eq.(\ref{3.7}).
Moreover, if $\theta=0$, $\lambda_{c}$ is given by

\begin{equation}
\label{3.9}
\lambda_{c}=\frac{3e^{3}}{\pi|\sigma|}.
\end{equation}

Thus, it seems that in contrast to the case $d=4$ treated in the
previous section, here the one-loop result is not reproduced.
However, it is important to remember that the one-loop
computation of Coleman and Weinberg neglects terms
proportional to $\lambda_{R}^{2}$. It was argued that under the
plausible hypothesis that $\lambda_{R}{\sim}e^{4}$, terms
proportional to $e^{4}$ and $\lambda_{R}$ correspond to the
leading contributions. The non-perturbative approach used in the
previous section confirm this hypothesis. Therefore, the $d=4$
result does not agree exactly with the complete one-loop
result. In fact, it agrees with the Coleman-Weinberg expression
of the one-loop result, which neglects some few terms. The same
thing happens in $d=3$. In order to have some insight with
respect to the order of magnitudes neglected, let us consider
for simplicity the case $\eta=\theta=0$.
In our approach we have that the
quartic self-coupling is given by Eq.(\ref{3.9}). According to
the one-loop result, a term ${\sim}e^{9/2}$ is being
neglected.

\section{Discussion} \setcounter{equation}{0}

The results of the previous sections show that the
phenomenon of dimensional transmutation can be established
outside of the perturbative framework. The Coleman-Weinberg
one-loop result for massless scalar $(QED)_{4}$ can be
obtained non-perturbatively. It has been shown recently
that this
approach allows a description of the Coleman-Weinberg
mechanism free from Landau ghost singularities \cite{Malbouisson}.
The Landau ghost singularity is frequently associated to a
trivial behavior of the theory. The one-loop renormalization
group analysis shows that the running coupling constant
have this problem \cite{Coleman}. In fact, it diverges for
finite momenta. The result of section 2 is obviously
non-perturbative and this trouble does not occur. The Landau
ghost seems to be an artifact of perturbation theory.

The case of the Maxwell-Chern-Simons
scalar QED was treated using
the prescription of section 2. We found the phenomenon of
dimensional transmutation and thence symmetry breaking. Of
course, this result was obtained at the expense of the
solution to a gap equation. It is well known that mean field
theories are characterized by gap equations. Also, mean field
theories have the tendency to produce phase transitions.
Therefore, it is not surprising that a broken symmetry solution
result from our computations. We use the gap equations to
eliminate the dimensionless couplings in the previous sections.
What has been done is just like mean field theories. Note that
mean field theories are necessarily non-perturbative. However,
it is not rare mean field theories produce spurious results.
A classical example is the Landau-Ginsburg theory for the
Ising ferromagnet \cite{Landau}. It turns out that the
mean field result reproduce the correct critical indices
only for $d>4$. For $d<4$ its predictions are completely
wrong. In fact, it predicts that symmetry breaking does
occur in $d=1$ while it is known from the exact $d=1$
solution that spontaneous magnetization is absent.
The case $d=2$ have exact solution \cite{Onsager}
and the Landau-Ginsburg theory disagrees with this exact result
although we have spontaneous magnetization in this case.
However, the Landau-Ginsburg theory of superconductivity gives
very good phenomenological results. This is because the
microscopic theory of superconductivity is itself a mean
field theory, the BCS theory \cite{Bardeen}. Indeed, the
Landau-Ginsburg theory of superconductivity can be
derived from the BCS theory \cite{Abrikosov}. The Landau-Ginsburg
theory of superconductivity is just the Higgs mechanism applied
to scalar QED. We have discussed another path to this theory
via the Coleman-Weinberg mechanism. We give to the
Coleman-Weinberg mechanism the status of mean field theory. The
Maxwell-Chern-Simons scalar QED may be viewed as a
phenomenological
theory to high temperature superconductors \cite{Fradkin}. It can
be shown that this system exibits also vortex solutions
\cite{Hong}.

Another path to the study of phase transitions is approach the
theory by finite temperature field theoretical methods
\cite{Dolan}. Recently the finite temperature technique has
been applied to the study of the Maxwell-Chern-Simons scalar
QED \cite{Toms}. It was found that a phase transition does
in fact occur for an infinitesimally small positive mass.
Therefore,
it seems that the transition is of the Coleman-Weinberg type.
It has been concluded in ref.\cite{Toms} that it is incorrect
to assume that the scalar field have an imaginary mass in
order to describe symmetry breaking in this system. This is a
characteristic of the $d=3$ case. It is more consistent to
break the symmetry in $d=3$ via the Coleman-Weinberg mechanism
than that with the Higgs machanism which needs an imaginary
mass introduced by hand.

Summarizing, we have obtained the phenomenon of dimensional
transmutation in Maxwell-Chern-Simons scalar QED. In addition,
we obtained that the vacuum is more stable for $\theta\neq 0$
than that for $\theta=0$. It is important to stress that the
approach used in this paper is non-perturbative. Indeed, it is
a tree level analysis of an effective theory obtained through
an exact integration of the gauge fields. The finite temperature
case is under investigation. An important question
concerns the order of the phase transition.

\section*{Acknowledgment}

We would like to thanks Prof.A.P.C.Malbouisson for
valuable discussions. This work was supported by
Conselho Nacional de desenvolvimento
Cientifico e Tecnologico-CNPq.

\end{document}